\pdfoutput=1

\documentclass[
 amsmath,amssymb,
 reprint,%
]{revtex4-1}

\usepackage{graphicx}
\usepackage{dcolumn}
\usepackage{bm}
\usepackage[utf8]{inputenc}
\usepackage[T1]{fontenc}
\usepackage{mathptmx}
\usepackage{textcomp}

\begin{document}

\preprint{AIP/123-QED}

\title[Magnetic properties of Co/Ni-based multilayers with Pd and Pt insertion layers ]{Magnetic properties of Co/Ni-based multilayers with Pd and Pt insertion layers }

\affiliation{Institute of Physics, University of Augsburg, D-86135 Augsburg, Germany}

\author{M. Heigl}
 \email{michael.heigl@physik.uni-augsburg.de.}

\author{R. Wendler}
\author{S. Haugg}
\author{M. Albrecht}

\date{\today}

\begin{abstract}
In this study, the influence of Pd and Pt insertion layers in Co/Ni multilayers (MLs) on their magnetic properties, e.g. magnetic anisotropies, saturation magnetization, coercivity, magnetic domain size, and Curie temperature, is investigated. We compare three series of [Co/Ni/X]$_N$ ML systems (X = Pd, Pt, no insertion layer), varying the individual Co layer thickness as well as the repetition number $N$. All three systems behave very similarly for the different Co layer thicknesses. For all systems, a maximum effective magnetic anisotropy was achieved for MLs with a Co layer thickness between 0.15\,nm and 0.25\,nm. The transition from an out-of-plane to an in-plane system occurs at about 0.4\,nm of Co. While [Co(0.2\,nm)/Ni(0.4\,nm)]$_N$ MLs change their preferred easy magnetization axis from out-of-plane to in-plane after 6 bilayer repetitions, insertion of Pd and Pt results in an extension of this transition beyond 15 repetitions. The maximum effective magnetic anisotropy was more than doubled from 105\,kJ/m$^3$ for [Co/Ni]$_3$ to 275 and 186\,kJ/m$^3$ for Pt and Pd, respectively. Furthermore, the insertion layers strongly reduce the initial saturation magnetization of 1100\,kA/m of Co/Ni MLs and lower the Curie temperature from 720 to around 500\,K.
\end{abstract}

\maketitle

\section{\label{sec:Intro}Introduction}

Magnetic thin film systems with perpendicular magnetic anisotropy (PMA) have been and are still intensively investigated. PMA is displayed in several Co-based multilayered structures \cite{Hashimoto1989,Etz2008,Girod2009,Gottwald2012,Beaujour2007,DenBroeder1991,Daalderop1992,Arora2017} and was shown to be useful for a variety of applications, including perpendicular magneto-optical recording \cite{Materials1990,Fullerton2014}, spin-transfer-torque magneto-resistive random access memories (STT-MRAM)\cite{Zhang2014,Katine2008,Meena2014,Sankey2008,L.YouR.C.SousaS.BandieraB.Rodmacq2012}, domain-wall-motion-based devices \cite{Yang2015, LeGall2015}, bit patterned media \cite{Albrecht2015,Albrecht2013,Albrecht2005, Hellwig2009}, and biomedical applications \cite{Varvaro2019}. In this regard, Co/Ni multilayers (MLs) are of particular interest because of their additional high spin polarization \cite{Gimbert2011,Andrieu2018} and low intrinsic Gilbert damping \cite{Zhang2014,Beaujour2007,Song2013,Chen2008,Ertl1992,Seki2017}. PMA in these MLs is mainly contributed to interface magnetic anisotropy \cite{DenBroeder1991,Daalderop1992,Vojta1996,Arora2017} and magnetoelastic anisotropy \cite{Sander2004,Sakamaki2012,Chang1993}. Theoretical calculations predict a maximum PMA for Co/Ni MLs in the fcc(111) structure at a Co thickness of one monolayer and a Ni thickness of two monolayers\cite{G.H.O.DaalderopP.J.Kelly1990,Daalderop1992} ($\approx$ Co(0.2\,nm)/Ni(0.4\,nm)), which has been experimentally confirmed by multiple groups \cite{F.J.A.denBroederE.Janssen1992, Bloemen1992, Gottwald2012,Girod2009,Carcia1988,Knepper2005}. Previous studies have shown that PMA of Co/Ni MLs is strongly affected by the deposition process \cite{Akbulut2015,Posth2009}, choice of seed layers\cite{Liu2017}, and post-treatment processing, e.g. annealing \cite{Zhang1995,Kurt2010} or ion radiation \cite{Stanescu2008,Sakamaki2012}. This limits the feasibility for many applications where the seed layer is not freely selectable or post-deposition treatments are no option. On the other hand, Co/Pt and Co/Pd MLs also exhibit large PMA and are less sensitive to seed layers or sputter-deposition conditions \cite{Schuppler2006,Ulbrich2008,Hashimoto1989,Carcia1988,Nakajima1998,Zeper1989}. 
 \\
 In this work, we combine these systems in order to enhance PMA by adding Pt or Pd insertion layers to Co/Ni MLs. These trilayer-based films provide a highly tunable system for a  wide range of applications. A previous study of Ni/Co/Pt and Co/Ni/Pt MLs has already demonstrated enhanced PMA and annealing stability compared to Co/Ni MLs \cite{Chen2015}. It has also been shown that the stacking order of the trilayer, as well as the Pt thickness is important. Here, we focus on Pd insertion layers and investigate their impact on the magnetic properties as a function of ML repetition number and Co layer thicknesses and provide a comparison to Co/Ni and Co/Ni/Pt MLs.

\section{\label{sec:Exp}Experimental Details}

\begin{figure}
\includegraphics[width=1\linewidth]{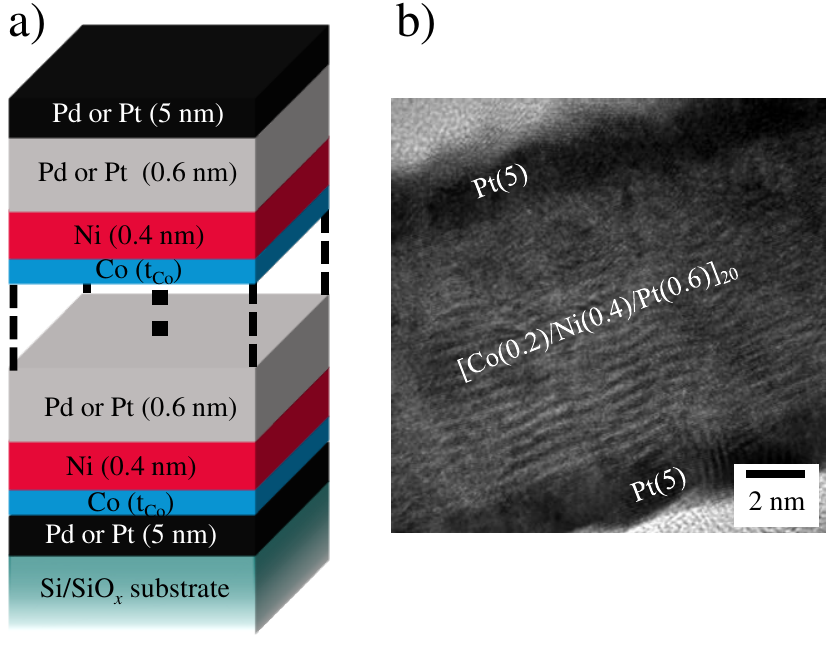}
\caption{\label{fig:stack} (a) Schematic image illustrating the layer stacking of Co/Ni/Pd(Pt) MLs. (b) Exemplary cross-section TEM image of a Pt(5\,nm)/[Co(0.2\,nm)/Ni(0.4\,nm)/Pt(0.6\,nm)]$_{20}$/Pt(5\,nm) film sample.}
\end{figure}

A series of [Co($t_{\mathrm{Co}}$)/Ni(0.4)/X(0.6)]$_N$ MLs (X = Pd, Pt, no insertion layer; thicknesses are given in nm) with various repetition numbers $N$ and different individual Co layer thicknesses $t_{\mathrm{Co}}$ were investigated. The films were deposited at room temperature by magnetron sputtering (base pressure < $10^{-8}$\,mbar) from elemental targets. The Ar pressure was kept constant at 5x$10^{-3}$\,mbar during the deposition process. The individual layer thicknesses were determined using a calibrated deposition rate. The films were prepared on Si(001) substrates with a 100\,nm thick thermally oxidized SiO$_x$ layer. 5\,nm of Pt were used as seed and capping layer for the Co/Ni and Co/Ni/Pt MLs. Accordingly, 5\,nm of Pd were used as seed and capping layer for the Co/Ni/Pd MLs. 
\\
We fixed the Ni and Pd/Pt layer thickness to 0.4 and 0.6\,nm, respectively, and varied the Co layer thicknesses between 0.1 and 0.5\,nm. The bi- and trilayers were repeated to form MLs with repetition numbers $N$ between 1 and 15.
A schematic of the layer stack is depicted in figure\,\ref{fig:stack}\,a) along with an exemplary transmission electron microscopy (TEM) cross-section image of a [Co(0.2)/Ni(0.4)/Pt(0.6)]$_{20}$ film sample (figure\,\ref{fig:stack}\,b), confirming the overall layer structure. Note that it is not possible to differentiate between Co and Ni layers because of their similar atomic number (Z contrast). Thus, only the contrast between two equally thick layers of Co/Ni and Pt is visible.
\\
The integral magnetic properties of the samples were investigated by superconducting quantum interference device-vibrating sample magnetometry (SQUID-VSM). $M$-$\mu_0H$ hysteresis loops were measured in out-of-plane (oop) and in-plane (ip) geometry at room temperature. 
It has been shown that edge effects can lead to artifacts in the measured $M$-$\mu_0H$ hysteresis loops \cite{Mandru2020}. Thus, their occurrence was prevented by cutting all edges of the measured samples. The effective magnetic anisotropy $K_{\mathrm{eff}}$ was determined from the integrated area difference between the oop and ip $M$-$\mu_0H$ hysteresis loops. Please note that for calculating the saturation magnetization $M_{\mathrm{s}}$, the total film volume (including the insertion layers) was taken into account. In order to obtain the Curie temperature, the oop or ip remanent magnetization was measured in the temperature range between 300 and 800\,K with a magnetic guiding field of 10\,mT. Magnetic force microscopy (MFM) was used to receive information about the magnetic domain structure after sample demagnetization at room temperature. In order to access the equilibrium domain size, the samples were demagnetized by decaying alternating magnetic fields.

\section{\label{sec:Results}Results and Discussion}

\subsection{\label{sec:Thickness}Co layer thickness dependence}

\begin{figure}
\includegraphics[width=1\linewidth]{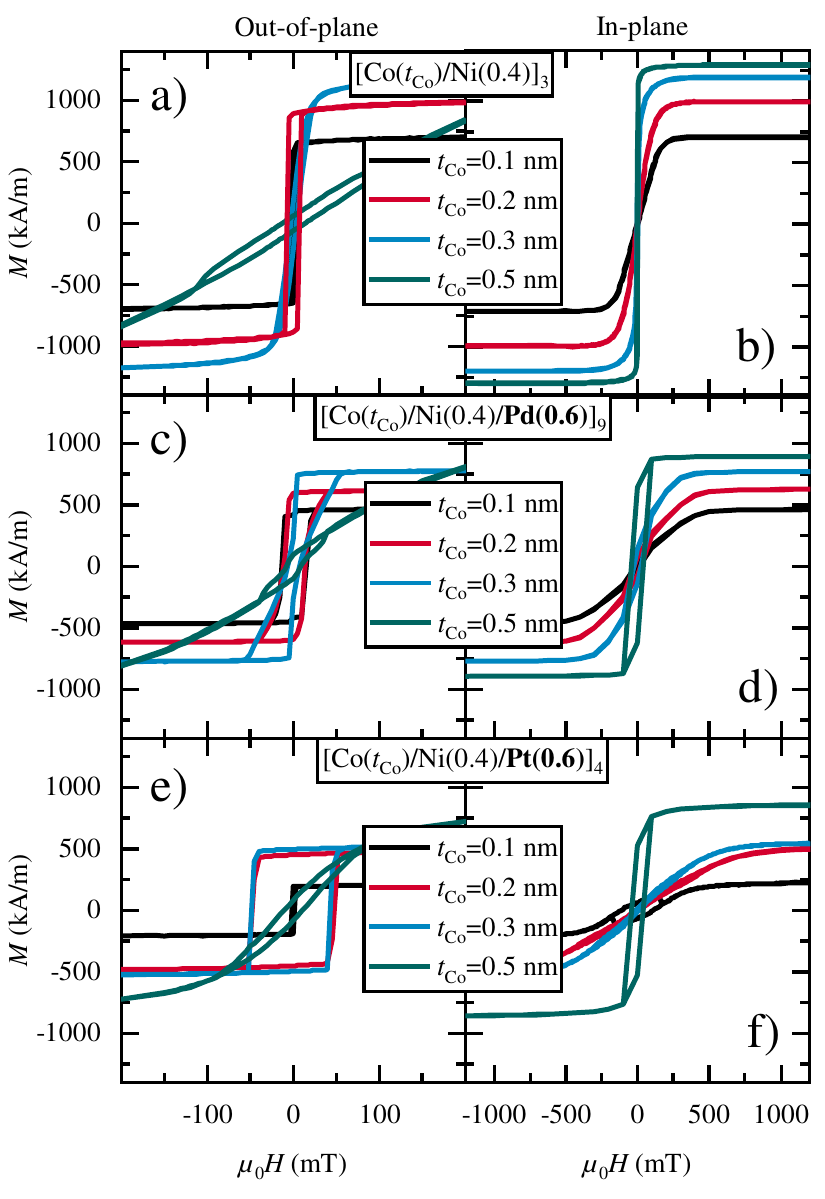}
\caption{\label{fig:LoopsCo} $M$-$\mu_0H$ hysteresis loops obtained in oop and ip geometry of [Co($t_{\mathrm{Co}}$)/Ni(0.4)]$_3$ (a, b), [Co($t_{\mathrm{Co}}$)/Ni(0.4)/Pd(0.6)]$_9$ (c, d), and [Co($t_{\mathrm{Co}}$)/Ni(0.4)/Pt(0.6)]$_4$ (e, f) with varying Co layer thickness. (All thicknesses are given in nm)}
\end{figure}

\begin{figure}
\includegraphics[width=1\linewidth]{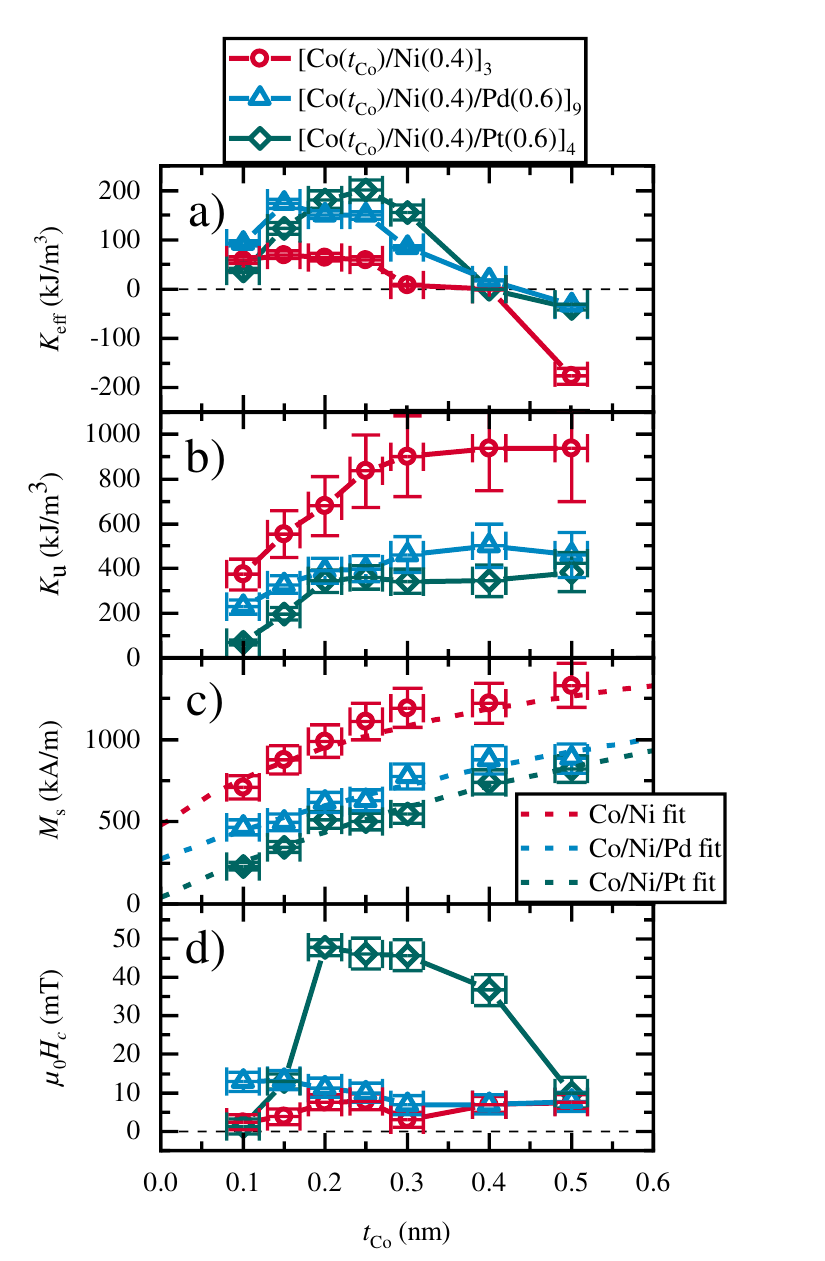}
\caption{\label{fig:ValuesCo} Effective magnetic anisotropy $K_{\mathrm{eff}}$ (a), uniaxial magnetic anisotropy $K_{\mathrm{u}}$ (b), saturation magnetization $M_{\mathrm{s}}$ (c), and coercivity field $\mu_0H_\mathrm{c}$ (d) as function of Co layer thickness $t_{\mathrm{Co}}$. (All thicknesses are given in nm)}
\end{figure}

 In a first study, the individual Co layer thickness $t_{\mathrm{Co}}$ was varied between 0.1 and 0.5\,nm for three sample series, [Co($t_{\mathrm{Co}}$)/Ni]$_3$, [Co($t_{\mathrm{Co}}$)/Ni/Pd]$_9$, and [Co($t_{\mathrm{Co}}$)/Ni/Pt]$_4$. The Ni and Pd/Pt thicknesses were set to 0.4 and 0.6\,nm, respectively. The repetition numbers were chosen to ensure an oop easy axis at thinner Co thicknesses. In figure\,\ref{fig:LoopsCo}, exemplary oop and ip $M$-$\mu_0H$ hysteresis loops for four different Co layer thicknesses are displayed. Based on the $M$-$\mu_0H$ data, the magnetic properties of the three sample series as a function of Co layer thickness were extracted and summarized in figure\,\ref{fig:ValuesCo}. 
 \\
 The effective magnetic anisotropy $K_{\mathrm{eff}}$ consists of the uniaxial magnetic anisotropy $K_{\mathrm{u}}$ and the magnetic shape anisotropy $K_{\mathrm{sh}}=\frac{\mu_0}{2} M_{\mathrm{s}}^2$:

\begin{eqnarray}
K_{\mathrm{eff}}=K_{\mathrm{u}}-\frac{\mu_0}{2} M_{\mathrm{s}}^2.
\label{eq:Ku}
\end{eqnarray}

The measured $K_{\mathrm{eff}}$ values are displayed in figure\,\ref{fig:ValuesCo}\,a). Positive values of $K_{\mathrm{eff}}$ imply an oop easy axis, negative values an ip easy axis. If the shape anisotropy is larger than the uniaxial one, a transition from an oop to an ip easy axis occurs. 
 The [Co($t_{\mathrm{Co}}$)/Ni(0.4)]$_3$ series shows a maximum $K_{\mathrm{eff}}$ of $70\pm7$\,kJ/m$^3$ for 0.15\,nm $\leq {t_{\mathrm{Co}}}$ < 0.20\,nm. For thicker Co layers, $K_{\mathrm{eff}}$ decreases and the easy axis direction changes from oop to ip between 0.30\,nm < ${t_{\mathrm{Co}}}$ < 0.40\,nm. 
 These results confirm previous studies on thickness dependencies in Co/Ni MLs revealing an optimal thickness ratio between Co and Ni of about one to two in order to get high PMA \cite{F.J.A.denBroederE.Janssen1992, Bloemen1992, Gottwald2012,Girod2009,Carcia1988,Knepper2005}. The sample series with insertion layers behave qualitatively similar. All systems have their maximum $K_{\mathrm{eff}}$ for a Co layer thickness between 0.15 and 0.25\,nm and their easy axis changes to the in-plane orientation at around 0.40\,nm. However, in comparison to Co/Ni, the maximum $K_{\mathrm{eff}}$ is more than doubled with insertion layers, where [Co(0.15)/Ni(0.4)/Pd(0.6)]$_9$ has a $K_{\mathrm{eff}}$ of $173\pm17$\,kJ/m$^3$ and [Co(0.25)/Ni(0.4)/Pt(0.6)]$_4$ of $201\pm20$\,kJ/m$^3$.
 This increase in anisotropy can be mainly attributed to the lower magnetization $M_{\mathrm{s}}$ (see figure\,\ref{fig:ValuesCo}\,c)) and thus a smaller $K_{\mathrm{sh}}$. Additionally, the insertion layer tend to prevent intermixing at the Ni and Co interface, which might give rise to an increased interface magnetic anisotropy \cite{Liu2017,LeGall2015,Bandiera2011}. 
 \\ 
Generally, $K_{\mathrm{u}}$ mainly arises from interface effects in these systems \cite{DenBroeder1991,Daalderop1992,Vojta1996,Arora2017}. Thus, $K_{\mathrm{u}}$ should not be thickness-dependent. This is only observable in figure\,\ref{fig:ValuesCo}\,b) for $t_{\mathrm{Co}}$ > 0.2\,nm, where $K_{\mathrm{u}}$ saturates for all systems. Below this thickness, the Co layer is particularly influenced by roughness and non-continuous growth. Co/Ni MLs show larger $K_{\mathrm{u}}$ values despite the lower interface anisotropy terms of Co/Ni (0.31\,mJ/m$^2$)\cite{Daalderop1992} in comparison to Co/Pt (0.50\,mJ/m$^2$)\cite{DenBroeder1991} and Co/Pd (0.40\,mJ/m$^2$)\cite{DenBroeder1991}. This can be explained by the higher number of interfaces per thickness. While Co(0.2)/Ni(0.4) MLs have 3 interfaces per 1.2\,nm of film, Co(0.2)/Ni(0.4)/X(0.6) MLs contain only 2. Additionally, the smaller amount of repetitions increases the impact of the interfaces to the Pt and Pd seed and capping layer.
\\
The saturation magnetization $M_{\mathrm{s}}$ of all systems displayed in figure\,\ref{fig:ValuesCo}\,c) increases with Co layer thickness. The behavior of [Co($t_{\mathrm{Co}}$)/Ni(0.4)]$_3$ can be modeled by a linear combination of two individual magnetization contributions:

\begin{eqnarray}
M_{\mathrm{s}}= \frac
{{t_{\mathrm{Co}}}M_{\mathrm{s,Co}}+0.4M_{\mathrm{s,Ni}}}
{{t_{\mathrm{Co}}}+0.4} 
\label{eq:Ku}
\end{eqnarray}

The magnetizations of the Co layer $M_{\mathrm{s,Co}}$ and Ni layer $M_{\mathrm{s,Ni}}$ were chosen as fit parameters. The measured data and fit can be seen in figure\,\ref{fig:ValuesCo} c).  $M_{\mathrm{s,Co}}$ has the fit value $1892$\,kA/m. It exceeds the bulk value of 1440\,kA/m \cite{coey_2010}, which can be explained by the higher magnetization of ultra-thin Co films \cite{Srivastava1998} and the increasing polarisation of the Pt seed layer with Co layer thickness. D. Weller et al. showed that there is no additional magnetic moment enhancement for Co/Ni interfaces. \cite{D.WellerY.WuJ.StohrandM.G.SamantB.D.Hermsmeier1994}. The fit value of $M_{\mathrm{s,Ni}}=480$\,kA/m agrees well with the bulk value of 488\,kA/m \cite{coey_2010}. 
The magnetization of the MLs with insertion layers X can be fitted in a similar way:

\begin{eqnarray}
M_{\mathrm{s}}= \frac
{{t_{\mathrm{Co}}}M_{\mathrm{s,Co+X}}+(0.4+0.6)M_{\mathrm{s,Ni+X}}}
{{t_{\mathrm{Co}}}+0.4+0.6} 
\label{eq:Ku}
\end{eqnarray}

For this fit, the polarization of the insertion layer is included  both in the $t_{\mathrm{Co}}$-dependent, as well as the $t_{\mathrm{Co}}$-independent term. 
The fit parameter $M_{\mathrm{s,Co+X}}$ for Co($t_{\mathrm{Co}}$)/Ni(0.4)/Pd(0.6) converges at $2231$\,kA/m. This increase in magnetization compared to bulk Co can mainly be explained by the polarization of Pd, which is dependent on $t_{\mathrm{Co}}$ in the analyzed thickness range\cite{Sander2004}. The $t_{\mathrm{Co}}$-independent fit parameter $M_{\mathrm{s,Ni+X}}$ for Co/Ni/Pd equals $272$\,kA/m. Beside the magnetization of the Ni layer, the Pd layer polarized by Co \cite{Sander2004} and Ni \cite{Chafai2011}   contributes. 
The measured values of [Co($t_{\mathrm{Co}}$)/Ni(0.4)/Pt(0.6)]$_4$ lead to the fit parameters $M_{\mathrm{s,Co+X}}=2413$\,kA/m and $M_{\mathrm{s,Ni+X}}=40$\,kA/m. The high value of $M_{\mathrm{s,Co+X}}$ is again due to the polarization of Pt by Co \cite{}. The interface between Ni and Pt is known to form a nonmagnetic NiPt alloy \cite{Benkirane2005}, which leads to an effectively lower magnetization for the Ni and Pt layers.
If we compare the two insertion layers magnetization-wise, it is important to note that the repetition number of [Co($t_{\mathrm{Co}}$)/Ni(0.4)/Pt(0.6)]$_4$ is lower than that of [Co($t_{\mathrm{Co}}$)/Ni(0.4)/Pd(0.6)]$_9$. Despite that, Pd generally shows a slightly larger induced magnetization which is less dependent on the Co layer thickness. The same observation was made in reference\cite{D.WellerY.WuJ.StohrandM.G.SamantB.D.Hermsmeier1994} where Co/Pd and Co/Pt MLs were compared. It was reported that Co/Pd MLS exhibit a magnetization value 15\% larger than that of Co/Pt MLs due to enhanced orbital moment of Co and larger polarization of Pd. 
\\
Figure\,\ref{fig:ValuesCo}\,d) shows the coercivity field $\mu_0H_\mathrm{c}$ of the oop hysteresis loops for the three sample series. The increase up to $t_{\mathrm{Co}}=$0.2\,nm can again be explained by the roughness of the sample and non-continuous growth of the Co layer. All sample series have their maximum at around 0.2\,nm with the highest value obtained for the [Co/Ni/Pt]$_4$ ML.

\subsection{\label{sec:Repetition}Multilayer repetition dependence}

\begin{figure}
\includegraphics[width=1\linewidth]{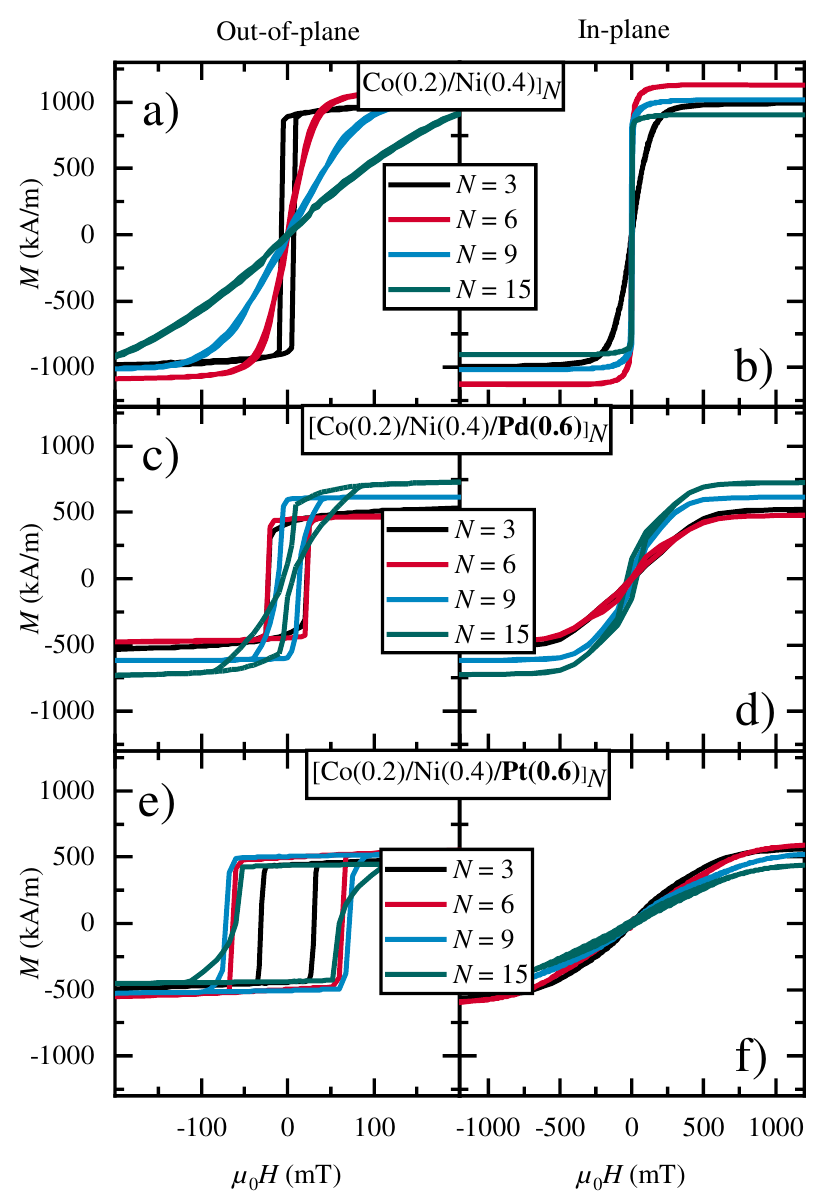}
\caption{\label{fig:LoopsN} $M$-$\mu_0H$ hysteresis loops obtained in oop and ip geometry of the three sample series with varying repetition number $N$. (All thicknesses are given in nm)}
\end{figure}

\begin{figure}
\includegraphics[width=1\linewidth]{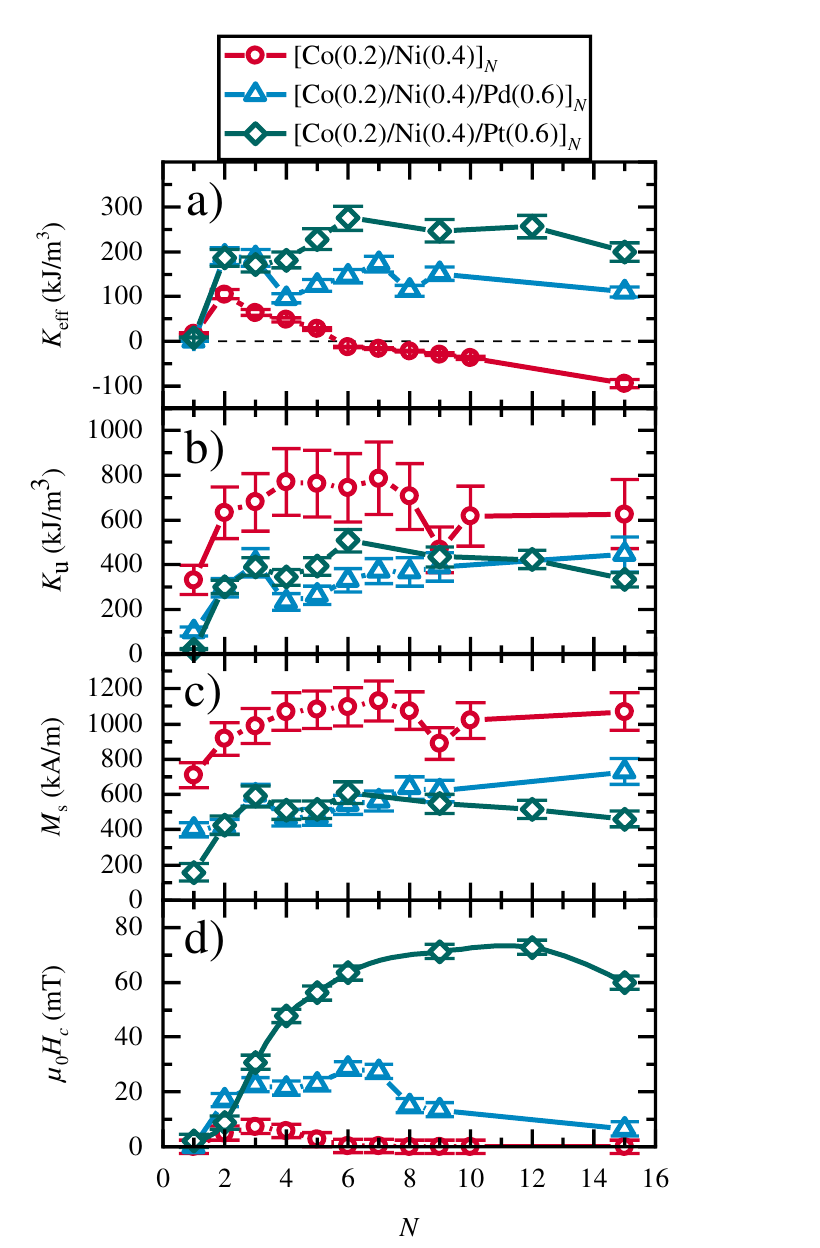}
\caption{\label{fig:ValuesN} Effective anisotropy $K_{\mathrm{eff}}$ (a), Uniaxial anisotropy $K_{\mathrm{u}}$ (b), saturation magnetization $M_{\mathrm{s}}$ (c), and coercivity field $\mu_0H_\mathrm{c}$ (d) as function of repetition number $N$. (All thicknesses are given in nm)}
\end{figure}

In a further study, we investigated the dependence on the repetition number $N$ for the following three sample series: [Co(0.2)/Ni(0.4)]$_N$, [Co(0.2)/Ni(0.4)/Pd(0.6)]$_N$, and [Co(0.2)/Ni(0.4)/Pt(0.6)]$_N$. Based on the results of the Co layer thickness study in section\,\ref{sec:Thickness}, $t_{\mathrm{Co}}=0.2$\,nm was selected for all films. The thicknesses of Ni and Pd/Pt stayed fixed at 0.4 and 0.6\,nm, respectively. 
Figure\,\ref{fig:LoopsN} shows exemplary oop and ip $M$-$\mu_0H$ hysteresis loops of the three sample series. Based on the $M$-$\mu_0H$ data, the magnetic properties of the three sample series as a function of repetition number were extracted and summarized in figure\,\ref{fig:ValuesN}. 
\\
In figure\,\ref{fig:ValuesN}\,a), the behavior of $K_{\mathrm{eff}}$ for $N\leq2$ can be explained by the onset of superparamagnetism in ultra-thin films \cite{Tang2006}. The values of $K_{\mathrm{eff}}$ in Co/Ni MLs change sign between $N=5$ and $N=6$ and with it their easy axes transition from oop to ip direction, which is mainly the result of increasing $M_{\mathrm{s}}$ and later of decreasing $K_\mathrm{u}$. There are two widely investigated approaches to delay this transition up to 10 repetitions: (i) increasing $K_\mathrm{u}$ by annealing \cite{Arora2017} and seed layer optimization \cite{Arora2017}, or (ii) decreasing $M_{\mathrm{s}}$ ($K_\mathrm{sh}$) by changing the ratio of Ni to Co \cite{Arora2017}. Introducing a paramagnetic insertion layer does both. The magnetization is reduced and the intermixing and roughness decreases \cite{Liu2017,LeGall2015,Bandiera2011}. This leads to a contrary behavior of $K_{\mathrm{eff}}$ in Co/Ni/X MLs: their $K_{\mathrm{eff}}$ values stay mostly constant within the studied repetition numbers. It was also possible to increase the maximum effective magnetic anisotropy more than twofold from $105\pm10\,\mathrm{kJ/m}^3$ for [Co/Ni]$_3$ to $186\pm19\,\mathrm{kJ/m}^3$ for [Co/Ni/Pd]$_3$ and $275\pm28\,\mathrm{kJ/m}^3$ for [Co/Ni/Pt]$_6$.  
\\
The $K_\mathrm{u}$ values shown in figure\,\ref{fig:ValuesN}\,b) increase for the Co/Ni MLs up to 4 repetitions to a maximum value of about 780$\pm$162\,kJ/m$^3$ and after 7 repetitions it starts to get reduced. The large error bars arise from the quadratic error propagation of $M_{\mathrm{s}}$. The behavior and values are in good agreement with recent work of Arora et al.\cite{Arora2017}. They showed that the moderate decrease of $K_\mathrm{u}$ for $N\geq10$ can be explained by increasing roughness and intermixing with larger $N$. The decrease for repetition numbers below 3 can be again explained by the onset of superparamagnetism. We observe a similar behavior for the sample series with insertion layers. The maximum $K_\mathrm{u}$ value of [Co/Ni/Pd]$_{15}$ and [Co/Ni/Pt]$_6$ amounts to 445$\pm$78\,kJ/m$^3$ and 507$\pm$51\,kJ/m$^3$, respectively, which is relatively small compared to 780$\pm$162\,kJ/m$^3$ of [Co/Ni]$_7$. The ratio of the $M_\mathrm{s}$ values between the three sample series shown in figure\,\ref{fig:ValuesN}\,c) stays more or less the same as already described in section\,\ref{sec:Thickness}. Moreover, we can observe superparamagnetic behavior in all three systems for low $N$, which is also apparent in the coercivity values $\mu_0H_\mathrm{c}$ (figure\,\ref{fig:ValuesN}\,d)).
\\

\begin{figure}
\includegraphics[width=1\linewidth]{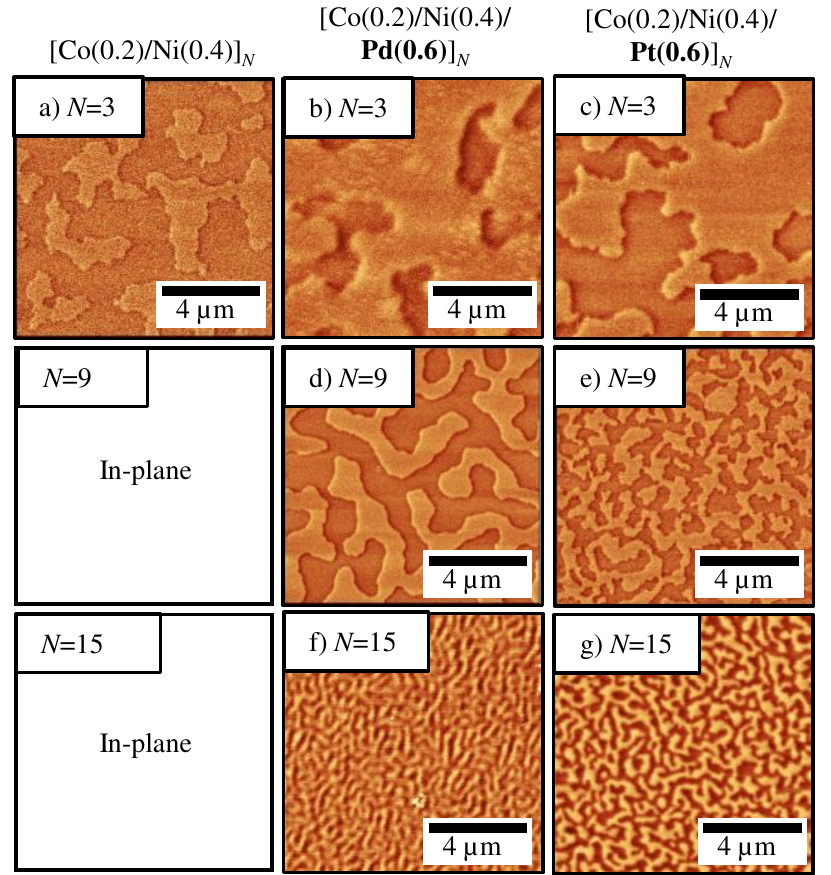}
\caption{\label{fig:MFM} Magnetic force microscopy images of [Co(0.2)/Ni(0.4)]$_N$ (a), [Co(0.2)/Ni(0.4)/Pd(0.6)]$_N$ (b, d, f), and [Co(0.2)/Ni(0.4)/Pt(0.6)]$_N$ (c, e, g) after sample demagnetization. (All thicknesses are given in nm)}
\end{figure}

In addition, we investigated the equilibrium domain size by MFM after demagnetizing the samples. Figure\,\ref{fig:MFM} shows MFM images of the three sample series with different repetition numbers. It was not possible to measure images of [Co(0.2)/Ni(0.4)]$_N$ for $N$=9 and 15 due to the ip orientation of the easy axis.
Theoretically, the magnetic domain size depends on the ratio $D_0$ of the domain wall and magnetostatic energies in the following way \cite{1993JMMM..128..111K}:

\begin{eqnarray}
D_0 = {\frac{4\pi(AK_\mathrm{u})^{1/2}}{\mu_0M_\mathrm{s}^2}} \;,
\label{eq:D0}
\end{eqnarray}

whereas $A$ is the magnetic exchange stiffness. The magnetic domain size $D$ can be estimated depending on the total film thickness $t$ by

\begin{eqnarray}
D(t) \varpropto t \cdot {\exp{\frac{D_0}{t}}} \; . 
\label{eq:D}
\end{eqnarray}
	
$D$ has its minimum at $t = D_0$ in equation\,\ref{eq:D}. For $t < D_0$, $D$ decreases rapidly with increasing $t$. [Co/Ni]$_3$ has the smallest $D_0$ of the analyzed samples. For Co/Ni MLs an exchange stiffness of $A\approx10^{-11}$\,J/m is commonly assumed\cite{Pellegren2017}. If we estimate $D_0$ with this value, we get $D_0\approx30$\,nm for [Co(0.2)/Ni(0.4)]$_3$. Thus, for all sample series the domain sizes decrease with higher repetition number (total thicknesses $t$). Please note that a direct comparison of the measured domain sizes with its theoretical estimation is hardly possible due to the exponential dependency of equation\,5 and the error range of the measured magnetic properties. Thus, in the following, the evolution of the domain sizes for the sample series is discussed only qualitatively. In the first row of figure\,\ref{fig:MFM}\,a-c) the domain sizes of MLs with $N=3$ are quite similar despite of the differences in measured magnetic properties between Co/Ni and Co/Ni/X (see figure\,\ref{fig:ValuesN}\,). The larger $K_\mathrm{u}$ and smaller $t$ values of [Co/Ni]$_3$ are offset by the larger $M_\mathrm{s}$ in equations 4 and 5. For larger repetition numbers, we can only compare the films with insertion layers. As already mentioned, the domain size is expected to decrease with increasing film thickness (for $t<D_0$), which is also observed in figure\,\ref{fig:MFM} b), d), f) for Co/Ni/Pd and c), e), g) for Co/Ni/Pt. A similar behavior was already reported for Co/Ni\cite{MacIa2012,AlRisi2020} and for Co/Pt\cite{Wang2020} MLs. Despite a smaller $K_\mathrm{u}$ and larger $M_\mathrm{s}$, [Co/Ni/Pd]$_9$ exhibits more sizeable domains, which could be a sign of a larger exchange stiffness $A$ in Co/Ni/Pd MLs. For 15 repetitions, [Co/Ni/Pt]$_{15}$ shows slightly larger domains, which can be explained by the smaller $M_\mathrm{s}$ value compared to [Co/Ni/Pd]$_{15}$.

\subsection{\label{sec:Curie}Curie temperatures}

\begin{figure}
\includegraphics[width=1\linewidth]{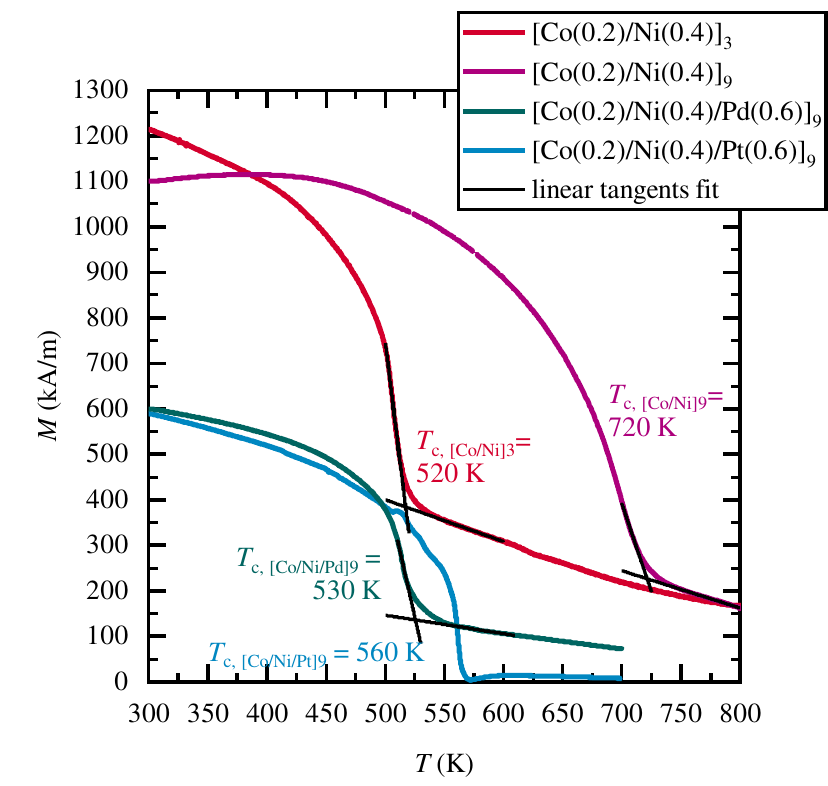}
\caption{\label{fig:Tc} Magnetization $M$ against temperature $T$ curve with a guiding field of 10\,mT. The Curie temperatures were estimated by linear fits around the turning point. (All thicknesses are given in nm) }
\end{figure}

Furthermore, we have investigated the Curie temperature $T_\mathrm{c}$ for selected Co/Ni/X MLs. In order to extract $T_\mathrm{c}$, the magnetization $M$ of [Co/Ni]$_3$, [Co/Ni]$_9$, [Co/Ni/Pd]$_9$, and [Co/Ni/Pt]$_9$ was measured dependent on the temperature $T$ (figure\,\ref{fig:Tc}). All systems were saturated at 300\,K and afterwards measured with increasing $T$ and an effective guiding field of 10\,mT. The measurement geometry of all samples was in oop direction, except [Co/Ni]$_9$, which was measured in ip geometry because of its ip easy axis. $T_{\mathrm{c}}$ was estimated by the intersecting tangents method.
\\
All $M-T$ curves, except for [Co/Ni/Pt]$_9$, do not drop fully to zero at their Curie temperature. Films measured oop might also lose magnetization due to decreasing $K_\mathrm{u}$, which is induced by interfacial intermixing at these elevated measurement temperatures\cite{Chen2015}. The Curie temperatures of [Co(0.2)/Ni(0.4)/Pd(0.6)]$_9$ and [Co(0.2)/Ni(0.4)/Pt(0.6)]$_9$ determined to 530\,K and 560\,K, respectively, are significantly lower than the $T_\mathrm{c}=720$\,K of the [Co(0.2)/Ni(0.4)]$_9$ ML without insertion layers. The reduction in exchange energy caused by the insertion layers is mostly responsible for that. If we compare [Co(0.2)/Ni(0.4)]$_3$ to the other samples, we find a $T_{\mathrm{c}}$ of 520\,K which is much lower than for the Co/Ni ML with 9 repetitions due to its smaller thickness. These values are in agreement with literature values. It was reported that $T_{\mathrm{c}}$ of Co/Pt and Co/Pd MLs lies between 520 and 570\,K \cite{F.J.A.denBroederH.W.vanKesterenW.Hoving1992,CARCIA1991} depending on the multilayer composition, while $T_{\mathrm{c}}$ of Co/Ni MLs can be as high as 950\,K, similar to Co$_{33}$Ni$_{67}$ alloys \cite{M.Hansen1958}.

\section{\label{sec:Conclusion}Conclusion}

The influence of Pd and Pt insertion layers on the magnetic properties of Co/Ni MLs was studied. It was shown that systems with and without insertion layers have a rather similar Co thickness dependence. An optimal Co layer thickness of $t_\mathrm{Co}=0.2$\,nm was validated for all systems in order to achieve strong PMA. While $K_\mathrm{eff}$ could be more than doubled, $M_\mathrm{s}$ and $K_\mathrm{u}$ decreased drastically with insertion of Pd and Pt.
The repetition number study revealed that $K_\mathrm{eff}$ could be increased over the whole measured range of repetition numbers of at least up to 15, allowing for extending the transition from an oop to an ip easy axis for larger repetition numbers for samples with insertion layers. MLs with Pt as insertion layer showed the largest $K_\mathrm{eff}$ and $\mu_0H_\mathrm{c}$. The variation of the magnetic domain sizes are consistent with the corresponding magnetic properties of the three sample series. Furthermore, insertion of Pd and Pt decreased the Curie temperature from 720\,K for [Co/Ni]$_9$ to 530\,K and 560\,K, respectively.

\begin{acknowledgments}
Financial support for this project was provided by the German Research Foundation (DFG, D.A.CH project AL 618/24-1) is gratefully acknowledged.
\end{acknowledgments}

\nocite{*}
\bibliography{CoNi-Pap}

\end{document}